\documentclass[twocolumn,superscriptaddress,pra]{revtex4-1}
\usepackage{mathrsfs}
\usepackage{amssymb, amsbsy, amsmath, latexsym, dsfont, array, layout, graphicx,mathrsfs,color}
\usepackage{soul}

\newcommand{\ket}[1]{\left|{#1}\right\rangle}
\newcommand{\bra}[1]{\left\langle{#1}\right|}

\begin{document}

\title{Experimental orthogonalization of highly overlapping quantum states with single photons}
\author{Gaoyan Zhu}
\affiliation{Department of Physics, Southeast University, Nanjing 211189, China}
\affiliation{Beijing Computational Science Research Center, Beijing 100084, China}
\author{Orsolya K\'{a}lm\'{a}n}
\affiliation{Institute for Solid State Physics and Optics, Wigner Research Centre,
Hungarian Academy of Sciences, P.O. Box 49, Hungary}
\author{Kunkun Wang}
\affiliation{Beijing Computational Science Research Center, Beijing 100084, China}
\author{Lei Xiao}
\affiliation{Department of Physics, Southeast University, Nanjing 211189, China}
\affiliation{Beijing Computational Science Research Center, Beijing 100084, China}
\author{Dengke Qu}
\affiliation{Department of Physics, Southeast University, Nanjing 211189, China}
\affiliation{Beijing Computational Science Research Center, Beijing 100084, China}
\author{Xiang Zhan}
\affiliation{School of Science, Nanjing University of Science and Technology, Nanjing 210094, China}
\author{Zhihao Bian}
\affiliation{School of Science, Jiangnan University, Wuxi 214122, China}
\author{Tam\'{a}s Kiss}
\affiliation{Institute for Solid State Physics and Optics, Wigner Research Centre,
Hungarian Academy of Sciences, P.O. Box 49, Hungary}
\author{Peng Xue}\email{gnep.eux@gmail.com}
\affiliation{Beijing Computational Science Research Center, Beijing 100084, China}

\begin{abstract}
We experimentally realize a nonlinear quantum protocol on single-photon qubits with linear optical elements and appropriate measurements. The quantum nonlinearity is induced by post-selecting the polarization qubit based on a measurement result obtained on the spatial degree of freedom of the single photon which plays the role of a second qubit. Initially, both qubits are prepared in the same quantum state and an appropriate two-qubit unitary transformation entangles them before the measurement on the spatial part. We analyze the result by quantum state tomography on the polarization degree of freedom. We then demonstrate the usefulness of the protocol for quantum state discrimination by iteratively applying it on either one of two slightly different quantum states which rapidly converge to different orthogonal states by the iterative dynamics. Our work opens the door to employ effective quantum nonlinear evolution for quantum information processing.
\end{abstract}

\maketitle

{\it Introduction:---}Quantum information processing protocols are known to exhibit speedup over classical algorithms due to specific features of quantum mechanics, such as linear superposition of quantum states or entanglement among subsystems. The usual assumption in quantum information theory is that the time evolution of the physical systems constituting the protocol is linear, e.g., in the case of a closed system the evolution is described by a unitary operator. If the constraint of linearity of the evolution is relieved, and a nonlinear equation governs the dynamics of the system, then one can design quantum protocols efficiently solving problems which are hard even for usual quantum algorithms \cite{AL98}. For example, the ability to quickly discriminate nonorthogonal states and thereby to solve unstructured search is a generic feature of nonlinear quantum mechanics \cite{CY16}. Nonlinear time evolution can be presented in standard quantum mechanics as an effective model, e.g., the Gross-Pitaevskii equation \cite{MW13} which approximately describes the collective behavior of atoms in a Bose-Einstein condensate. Were it not approximate, the Gross-Pitaevskii equation would be applicable to solve the unstructured search problem with an exponential improvement over protocols based on standard quantum theory \cite{CY16,LNT+18}.

An alternative way of introducing effective nonlinear evolution within the framework of standard quantum theory is to apply selective measurements in iterated protocols \cite{KJA+06}. The original idea of Bechmann-Pasquinucci, Huttner, and Gisin \cite{BPHG98} is based on the fact that if two identically prepared qubits are subjected to an entangling quantum operation, then by measuring one of the output qubits in one of the computational basis states $\ket{0}$, the quantum state of the other qubit undergoes an effective nonlinear transformation. The presence of two identical states at the input, together with the entangling transformation on the two qubits and the post-selection of the second qubit according to the result of the measurement on the first qubit, are the key elements leading to the emergent nonlinearity.


The resulting protocols, when applied iteratively, lead to highly nontrivial dynamics, with several intriguing features, such as a variety of fractals on the Bloch sphere representing the initial state of the qubit, leading to non-convergent, chaotic behavior~\cite{GKJ16,TBA+17,MJK+19}. One obviously cannot beat usual quantum efficiency limits in this way, since the emergent nonlinearity is an effective feature and one has to pay its cost in the form of discarded qubits \cite{GKJ16}, nevertheless, these protocols may find  applications for specific tasks, e.g. when matching a state to a reference with a prescribed maximum error \cite{KK18}.

The specific protocol we consider here is able to evolve any initial state to one of a pair of orthogonal states, according to a well-defined property of the initial state. Initial states, which have a positive $x$ coordinate on the Bloch sphere, will all converge to the quantum state pointing in the $+x$ direction, while the states with negative $x$ coordinate will converge to its orthogonal pair, the quantum state pointing in the $-x$ direction. Since the same protocol carries out this task for every initial state, one may demonstrate its effectiveness by comparing the convergence of highly overlapping initial states with $x$ components of opposite sign. Our protocol is thus able to discriminate any two quantum states with $x$ components of opposite sign unambiguously in the asymptotic limit. This approach is more general than standard optimal quantum state discrimination methods~\cite{I87,D88,P88,HMG+96,DB02,BCS+08}, where the discrimination of a pair of quantum states requires the construction of a specific protocol. After a finite number of steps, our protocol probabilistically enhances the overlap with one member of an orthogonal pair in a somewhat similar manner to the method proposed by Sol\'{i}s-Prosser et al.~\cite{SPD+16}
for the probabilistic separation of a finite number of quantum states.

Linear optics is a natural candidate among a variety of physical systems~\cite{LB99,ACM02} for realizing the protocols of quantum information processing~\cite{CN02}. In order to effectively implement quantum gates, linear optics has to be complemented by either optical elements exhibiting strong optical non-linearity~\cite{KB00} or, alternatively, apply post-selection with ancilla modes and projective measurements~\cite{TBA+17,GKJ16,KK18} resulting in probabilistic realizations.

In this paper, we realize the orthogonalization of quantum states via measurement-induced nonlinearity with single photons. We demonstrate that, after a few iterations of the nonlinear quantum transformation, one can substantially decrease the overlap of two, initially highly overlapping quantum states. After serval steps of the iterations they can become almost orthogonal to each other with only a small residual overlap.

%

\begin{figure}
\includegraphics[width=0.5\textwidth]{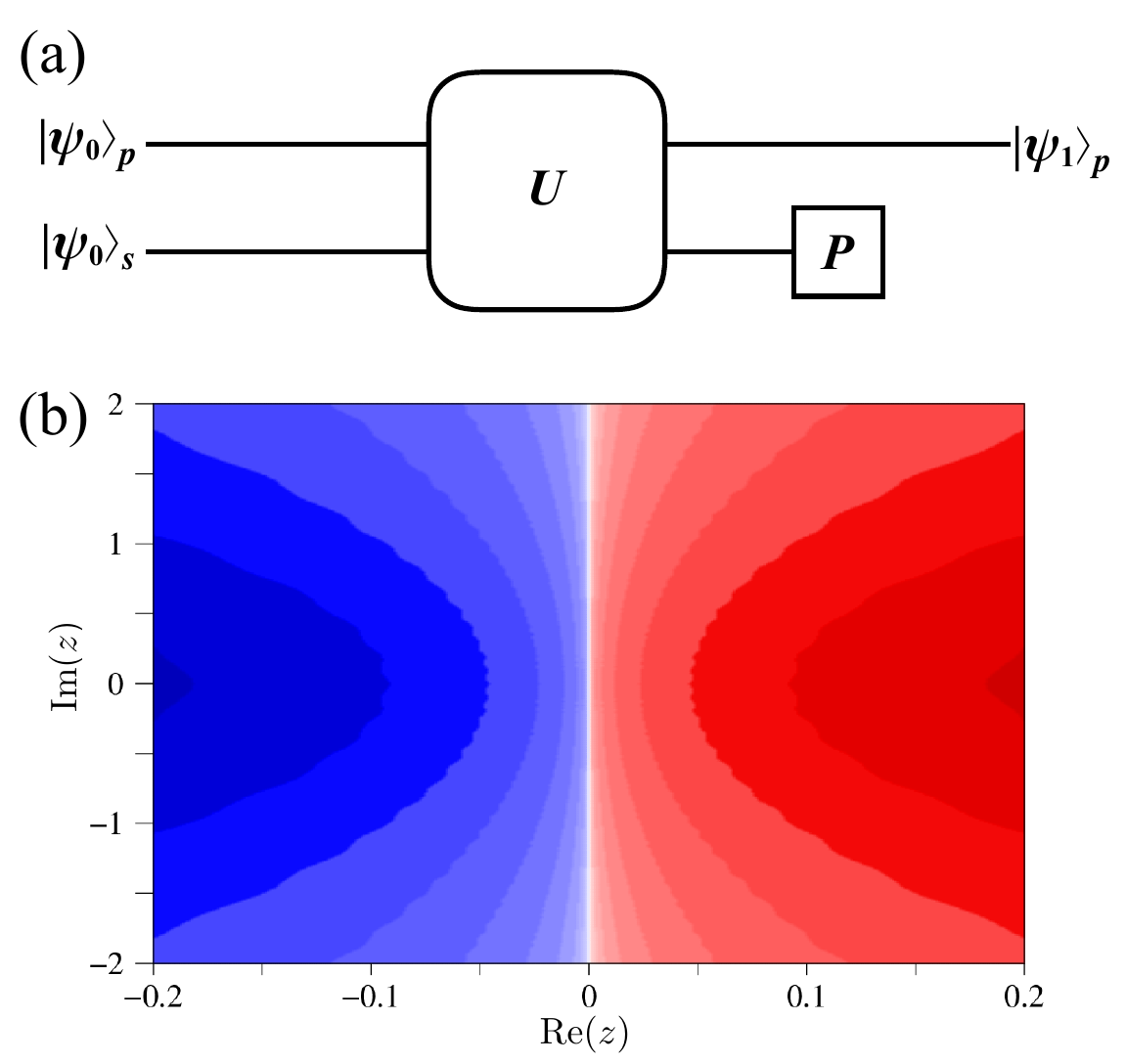}
\caption{(a) Schematic of one step of the nonlinear quantum protocol. $U$ and $P$ denote the entangling two-qubit transformation and the projective measurement, respectively. (b) The convergence regions of the corresponding complex map $f$ on the complex plane, where red (blue) color represents convergence to the asymptotic state $\ket{+}_{x}$ ($\ket{-}_{x}$), and the lighter the shading the more iterations are needed to reach the respective state. The white line represents the Julia set of the map.}
\label{fig:circuit}
\end{figure}

{\it Theoretical description of the protocol:---}Our aim is to implement a measurement-induced nonlinear quantum transformation~\cite{TBA+17} on photonic qubits. This can be realized on one member of a pair of qubits, initially in the same quantum state, via a controlled two-qubit unitary transformation on the composite system and a subsequent post-selective measurement on the other member of the pair (shown in Fig.~\ref{fig:circuit}(a)). For the two qubits, we consider two two-level systems: one encoded by the polarizations $\{\ket{H}=\ket{0}_{\text{p}}, \ket{V}=\ket{1}_{\text{p}}\}$ and the other by the spatial modes $\{\ket{D}=\ket{0}_{\text{s}},\ket{U}=\ket{1}_{\text{s}}\}$ of single photons.  Note that the subscripts $\text{p}$ and $\text{s}$ refer to the two types of degrees of freedom, respectively.

Initially, both qubits are prepared in the same quantum state $\ket{\psi_{0}}$, which can be described by the single complex parameter $z$, and the two-qubit system is thus a product state of the form
\begin{align}
\label{eq:initial}
\ket{\psi_0}_\text{p}\otimes\ket{\psi_0}_\text{s}=\frac{\ket{0}_{\text{p}}+z \ket{1}_{\text{p}}}{\sqrt{1+|z|^2}}\otimes\frac{\ket{0}_{\text{s}}+z \ket{1}_{\text{s}}}{\sqrt{1+|z|^2}}.
\end{align}
We apply the entangling two-qubit transformation
\begin{equation}
\label{eq:U}
U=\frac{1}{\sqrt{2}}\begin{pmatrix}
    1 & 0 & 0 & 1 \\
    0 & -1 & 1 & 0 \\
    0 &1 & 1 & 0 \\
    1 & 0 & 0 & -1 \\
    \end{pmatrix}
\end{equation}
after which the state of the composite system becomes
\begin{align} \ket{\Psi}_{\text{ps}}=\frac{1}{\sqrt{2}\left(1+\left|z\right|^{2}\right)}&\left[(1+z^2)\ket{00}_{\text{ps}}+2z\ket{10}_{\text{ps}}\right.\nonumber\\
&\left.+(1-z^2)\ket{11}_{\text{ps}}\right].
\label{2qubit_state}
\end{align}
Then, a projective measurement $P=\ket{D}\bra{D}=\ket{0}\bra{0}_{\text{s}}$ is applied on the spatial qubit by which one can post-select the polarization qubit in the state
\begin{equation}
\ket{\psi_1}_\text{p}=\frac{\ket{0}_{\text{p}}+f(z)\ket{1}_{\text{p}}}{\sqrt{1+|f(z)|^2}},
\end{equation}
where
\begin{equation}
f(z)=\frac{2z}{1+z^2}.
\label{f}
\end{equation}
The success probability of the first iteration of the protocol is dependent on the complex number $z$ characterizing the input state and can be formulated as
\begin{equation}
{\mathcal P}^{(1)}={\mathcal P}(z)=\frac{1}{2}+\frac{2\left(\mathrm{Re}z\right)^2}{\left(1+\left|z\right|^2\right)^2}.
\end{equation}
It can be seen that $\mathcal P^{(1)} \geq1/2$, the equality holds for $\mathrm{Re}z=0$, i.e., for the imaginary axis. In order to iterate the protocol, one needs to prepare also the spatial mode in state $\ket{\psi_1}_\text{s}$ is for the next step.

\begin{figure*}
\includegraphics[width=\textwidth]{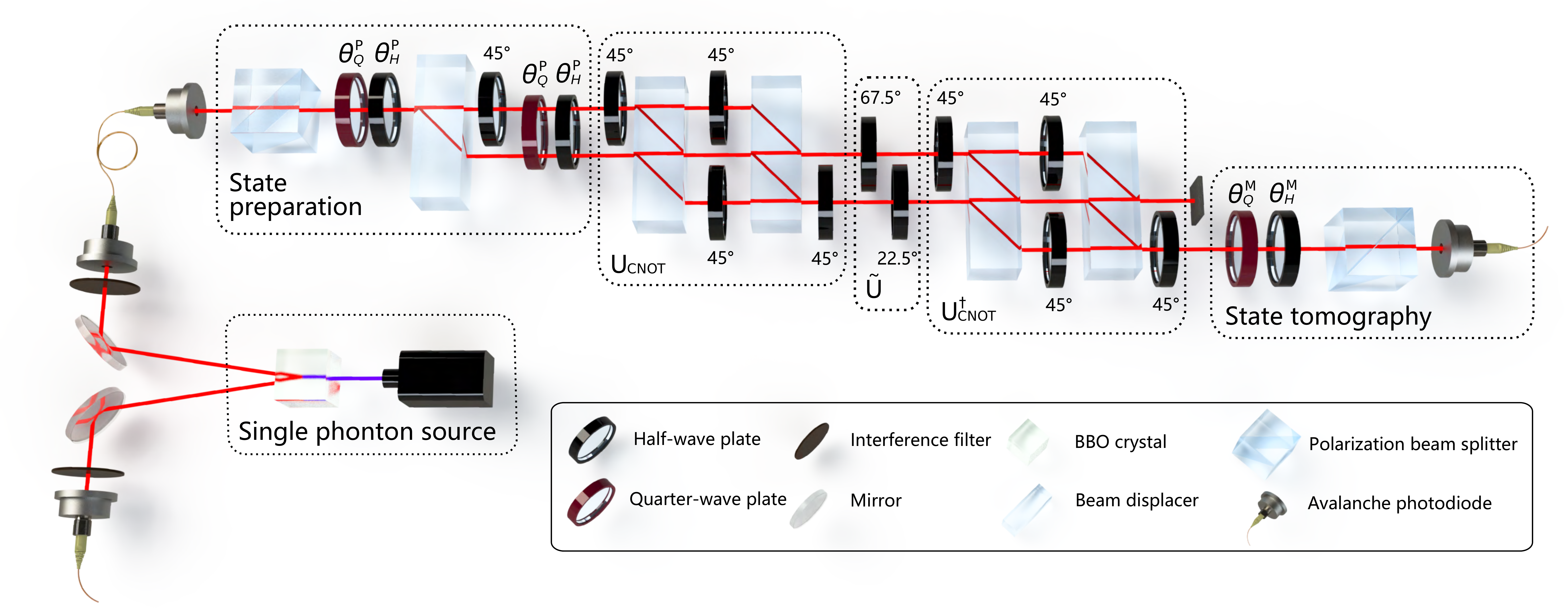}
\caption{
Experimental setup. Photon pairs are generated via type-I spontaneous parametric down conversion (SPDC). The pump is filtered out by an interference filter which restricts the photon bandwidth to $3$nm. With the detection of the trigger via avalanche photodiode, the heralded single photon is injected into the optical network involving three stages: state preparation, measurement-induced nonlinear transformation and state tomography. Transformation is realized with combination of beam displacers (BDs) and half-wave plates (HWPs) at certain angles. A projection is applied on the auxiliary qubit and the target qubit is polarization analyzed using a quantum state tomography system consisting of a HWP and a quarter-wave plate (QWP) followed by a polarizing beam splitter (PBS) in front of avalanche photodiode (APD). Trigger-herald pair is counted by the coincidence of APDs.}
\label{fig:setup}
\end{figure*}

In general, after $n$ iterations, the final state of the polarization qubit is $\ket{\psi_n}_\text{p}=(\ket{0}_{\text{p}}+f^{(n)}(z)\ket{1}_{\text{p}})/(\sqrt{1+|f^{(n)}(z)|^2})$, where $f^{(n)}(z)$ is defined recursively $f^{(n)}(z)=f[f^{(n-1)}(z)]$. The success probabilities of the second and the $n$th iterations are respectively $\mathcal{P}^{(2)}=\mathcal{P}\left[f(z)\right]$ and $\mathcal{P}^{(n)}=\mathcal{P}\left[f^{(n-1)}(z)\right]=1/2+2\{\mathrm{Re}\left[f^{(n-1)}(z)\right]\}^2/\left[1+|f^{(n-1)}(z)|^2\right]^2$. The success probability of orthogonalization -- or more precisely, of reaching an asymptotic state with a given precision, starting from an ensemble of qubits in the same initial state -- is a product of the single-iteration success probabilities $\prod_n \mathcal{P}^{(n)}$. We note that our setup is designed in a way that the projective measurement on the spatial qubit is automatically realized together with the post-selection whenever the photon is detected in the lower spatial mode (and not detected in the upper mode), see Fig.~\ref{fig:setup}.

The nonlinear transformation $f$ of Eq.~(\ref{f}) is a complex quadratic rational map \cite{M06,ML93}, which has been analyzed in~\cite{TBA+17}. It has two superattractive fixed points: $z_{1}=1$, and $z_{2}=-1$. Superattractiveness, which is related to the fact that $\left.\frac{df}{dz}\right|_{z_{i}}$=0 $(i=1,2)$, ensures that the convergence to the two fixed points $z_{1}$ and $z_{2}$ is fast. There is a set of points which do not converge to any of the attractive fixed points when iterating the map $f$ and these form the so-called Julia set of the complex map (the third fixed point of the map $z_{3}=0$, which is repelling, is also a member of the Julia set). The Julia set of the map $f$ is the imaginary axis on the complex plane (see Fig.~\ref{fig:circuit}(b)) or equivalently, the great circle which intersects the $y$ axis on the Bloch sphere, while the two superattractive fixed points correspond to the orthogonal quantum states
\begin{equation}
\ket{\psi_{z_{1}}}=\ket{+}_{x}=\frac{\ket{0}+\ket{1}}{\sqrt{2}}, \text{   }
\ket{\psi_{z_{2}}}=\ket{-}_{x}=\frac{\ket{0}-\ket{1}}{\sqrt{2}},
\end{equation}
pointing in the $+x$ and $-x$ directions on the Bloch sphere, respectively. It can be seen in Fig.~\ref{fig:circuit}(b) that initial states which can be described by a complex number $z$ that has a positive (negative) real part, all converge to the asymptotic state $\ket{+}_{x}$ ($\ket{-}_{x}$), as represented by the coloring. Initial states which lie closer to the border of these convergence regions (i.e., the Julia set) need more iterations to approach the respective asymptotic state.  It has been shown that by iterating the above procedure on two ensembles of qubits, the states of which initially have a large overlap, but have an $x$ component of opposite sign, then already a few iterational steps are enough to approximately orthogonalize them, thereby effectively implementing quantum state discrimination {\cite{TBA+17}}. Moreover, the scheme is applicable to sort all quantum states according to which part of the Bloch sphere they are initially from, without needing to modify the setup itself. Let us further note that the success probability of subsequent steps grows and approaches $1$ as the states converge to either of the asymptotic states.

In our experiment, it is always the polarization qubit which is kept after the post-selection and analyzed afterwards, while in every subsequent step both the states of the spatial qubit and the polarization qubit are reprepared according to the quantum state tomographic measurements performed on the polarization qubit in the previous step.

{\it Experimental realization:---}For experimental demonstration, pairs of photons are generated via type-I spontaneous parametric down-conversion (SPDC)~\cite{XZQ+15,BLQ+15,ZZL+16,ZKK+17,ZCL+17,WEX+18,WWZ+18}. With the detection of trigger photons, the other photons in one pair are heralded and act as a single photon source in the experimental setup shown in Fig.~\ref{fig:setup}. Experimentally, photon pairs are counted by coincidences between two single-photon avalanche photodiodes (APDs) with a time window of $3$ns. Total coincidence counts are about $12,000$ over a collection time of $3$s.

The heralded single photons pass through a polarizing beam splitter (PBS) followed by a quarter-wave plate (QWP) and a half-wave plate (HWP) with setting angles $\theta^\text{P}_Q$ and $\theta^\text{P}_H$, respectively. Then a birefringent calcite beam displacer (BD) splits them into two parallel spatial modes, i.e., upper and lower modes, depending on their polarizations. Photons in the upper mode pass through a HWP at $45^\circ$ to flip their polarizations from $\ket{V}$ to $\ket{H}$. Photons in both spatial modes pass through a QWP and a HWP with the setting angles $\theta^\text{P}_Q$ and $\theta^\text{P}_H$, respectively, and then they are prepared in the initial state (\ref{eq:initial}) with
\begin{equation}
z=\frac{i\sin2\theta^\text{P}_H+\sin(2\theta^\text{P}_H-2\theta^\text{P}_Q)}{i\cos2\theta^\text{P}_H+\cos(2\theta^\text{P}_H-2\theta^\text{P}_Q)}.
\end{equation}
Note that the matrix form of the operation of a HWP with setting angle $\theta$ reads
$\begin{pmatrix}
    \cos 2\theta & \sin 2\theta \\
    \sin 2\theta & -\cos 2\theta \\
    \end{pmatrix}$, and that of a QWP at $\vartheta$ reads
$\begin{pmatrix}
\cos^2 \vartheta+i \sin^2 \vartheta & (1-i)\sin\vartheta\cos\vartheta\\
(1-i)\sin\vartheta\cos\vartheta & \sin^2\vartheta+i\cos^2\vartheta \\ \end{pmatrix}$.

The unitary operation $U$ of Eq.~(\ref{eq:U}) is implemented as
\begin{align}
U=U^\dagger_\text{CNOT}\,\tilde{U}\,U_\text{CNOT},
\end{align}
where
\begin{align*}
\tilde{U}=U_\text{CNOT}UU^\dagger_\text{CNOT}=\frac{1}{\sqrt{2}}\begin{pmatrix}
    1 & 0 & 1 & 0 \\
    0 & -1 & 0 & 1\\
    1 & 0 & -1 & 0 \\
    0 & 1 & 0 & 1 \\
    \end{pmatrix},
\end{align*}
\begin{align*}
U_\text{CNOT}=U^\dagger_\text{CNOT}=\begin{pmatrix}
    1 & 0 & 0 & 0 \\
    0 & 1 & 0 & 0 \\
    0 & 0 & 0 & 1 \\
    0 & 0 & 1 & 0 \\
    \end{pmatrix}.
\end{align*}

Here we used the fact that the operation $U$ can be decomposed into operations $\tilde{U}$, and controlled-Not operation $U_\text{CNOT}$. Both of these unitary operations are controlled two-qubit rotations. $\tilde{U}$ can be realized by two HWPs, one at $67.5^\circ$ inserted into the upper mode and one at $22.5^\circ$ inserted into the lower mode. $U_{\text{CNOT}}$ can be realized by HWPs at $45^\circ$ and BDs. BDs are used to split the photons with different polarizations into different spatial modes and then to combine the two polarization modes into the same spatial mode. Then two-mode transformations can be implemented via HWPs acting on the two polarization modes propagating in the same spatial mode. For $U_{\text{CNOT}}$, HWPs at $45^\circ$ are used to flip polarizations of the photons.

\begin{figure*}
\includegraphics[width=\textwidth]{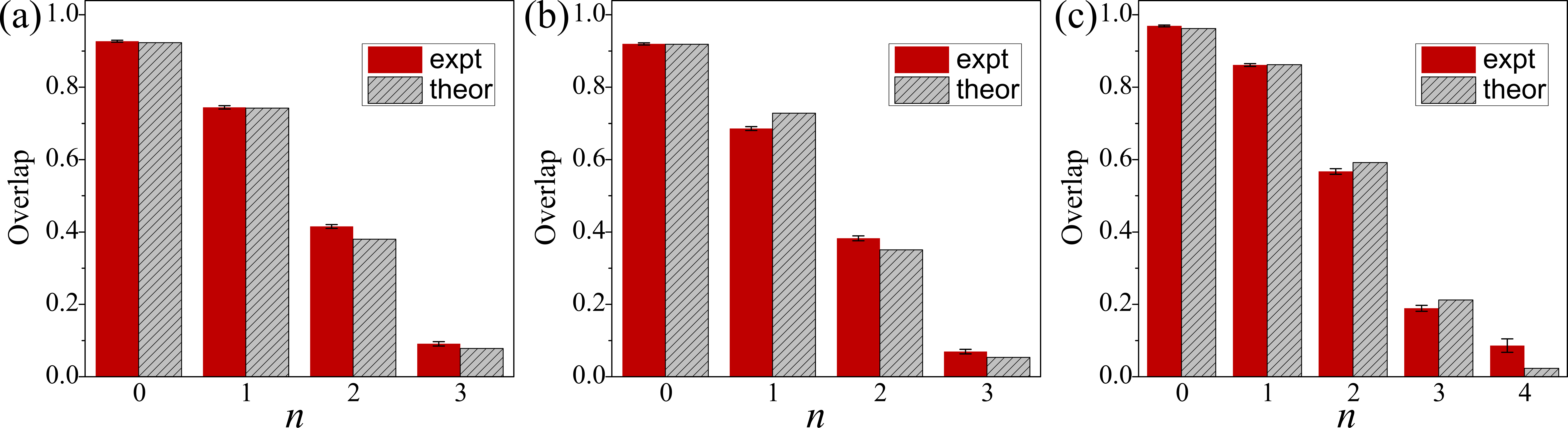}
\caption{
Overlap between three different pairs of quantum states in every iteration of the nonlinear transformation up to three (or four). Experimentally data obtained via quantum state tomography are shown by red color, and theoretical predictions are represented by grey bars. In the case of (a) the initial states for the protocol are described by the complex numbers $\{z_1=0.2,z_2=-0.2\}$, in (b) $\{z_1=0.2,z_2=-0.2-0.1i\}$, while in (c) $\{z_1=0.2e^{i\frac{\pi}{4}},z_2=-0.2e^{-i\frac{\pi}{4}}\}$. The statistical uncertainty is indicated by error bars which are calculated from Monte-Carlo simulations assuming Poissonian photon-counting statistics.}
\label{fig:overlap}
\end{figure*}

The post-selection of the polarization state can be realized by projecting the spatial qubit onto the basis state corresponding to the lower spatial mode $\ket{0}_{\text{s}}=\ket{D}$, where the polarization state of the photon is also analyzed. If a photon is detected in the upper spatial mode, then the nonlinear transformation of the polarization state does not take place (see Eq.~(\ref{2qubit_state})).

To demonstrate that the nonlinear protocol effectively orthogonalizes initially close quantum states~\cite{LZB+17,ZXB+17,WXQ+18,XQW+18,nc19,prl19}, the step presented in Fig.~\ref{fig:circuit}(a) has to be iterated, i.e., the initial state of the input qubits of the second step has to be equal to the output state $\ket{\psi_1}$ of the first step. In order to implement this, we use quantum state tomography to determine the output state after each step via a PBS, a QWP and a HWP with the setting angles $\theta^\text{M}_Q$ and $\theta^\text{M}_H$, respectively, projecting the output state into one of four different basis states $\{\ket{H},\ket{V},(\ket{H}+\ket{V})/\sqrt{2},(\ket{H}-i\ket{V})/\sqrt{2}\}$ to obtain the density matrix of the output state via maximum likelihood method. The resulting photons
are detected by APDs, in coincidence with the trigger photons. With the measured density matrices we reconstruct a pure state $\ket{\psi_1}$ with the method of minimum squares, which we prepare as initial state for both the polarization and the spatial qubit for the next iteration. Subsequent iterations are realized in the same way.


In our experiment, we chose three pairs of initial states $\ket{\psi_0(z_1)}$ and $\ket{\psi'_0(z_2)}$ to be discriminated by the nonlinear protocol. In Fig.~\ref{fig:overlap}, we show the experimental (red) and theoretical (grey) results of the overlaps for each iteration up to three (or four), starting from three different pairs of initial states. It can be seen that for the first pair of states (Fig.~\ref{fig:overlap}(a)), the overlap decreases from $0.927\pm0.003$ (the corresponding theoretical prediction is $0.923$) to $0.091\pm0.006$ (the corresponding theoretical prediction is $0.078$) after three iterations. For the second pair of states (Fig.~\ref{fig:overlap}(b)), the overlap decreases from $0.920\pm0.003$ (the corresponding theoretical prediction is $0.919$) to $0.070\pm0.006$ (the corresponding theoretical prediction is $0.054$) after three iterations. For the third pair of states (Fig.~\ref{fig:overlap}(c)), the overlap decreases from $0.969\pm0.002$ (the corresponding theoretical prediction is $0.962$) to $0.086\pm0.019$ (the corresponding theoretical prediction is $0.023$) after four iterations. Our experimental results agree well with those of the theoretical model, and the slight difference between the experimental data and theoretical values is due to the imperfections of the experiment. The results prove that the nonlinear transformation orthogonalizes the states in a few iterations and can therefore be employed for discriminating quantum states.

{\it Summary:---}We experimentally generated measurement-induced nonlinear transformations by linear optical elements and post-selective measurements on qubits represented by single photons. We demonstrated that such a transformation, experimentally realized for the first time, can be applied for the approximate orthogonalization of states with high initial overlap. Via the orthogonalization procedure we can prepare the qubits in distinguishable states so that they can be either directly measured or used for further processing. This measurement-induced nonlinear evolution can be considered as an implementation of a Schr\"{o}dinger microscope~\cite{GKJ16,LS00}. In a more general context a similar protocol can be applied for quantum state matching~\cite{KK18}.

\acknowledgements
This work has been supported by the National Natural Science Foundation of China (Grant Nos. 11674056 and U1930402), and the startup funding of Beijing Computational Science Research Center. T. K. and O. K. are grateful for the support of the National Research, Development and Innovation Office of Hungary (Project Nos. K115624, K124351, PD120975, 2017-1.2.1-NKP-2017-00001).


\begin{references}
\bibitem{AL98} D. S. Abrams and S. Lloyd, Phys. Rev. Lett. {\bf 81}, 3992 (1998).
\bibitem{CY16} A. M. Childs and J. Young, Phys. Rev. A {\bf 93}, 022314 (2016).
\bibitem{MW13} D. A. Meyer and T. G. Wong, New J. Phys. {\bf 15}, 063014 (2013).
\bibitem{LNT+18} K. de Lacy, L. Noakes, J. Twamley and J. B. Wang, Quantum Inf. Process. {\bf 17}, 266 (2018).
\bibitem{KJA+06}  T. Kiss, I. Jex, G. Alber and  S. Vym\v etal, Phys. Rev. A {\bf 74}, 040301(R) (2006).
\bibitem{BPHG98} H. Bechmann-Pasquinucci, B. Huttner and N. Gisin, Phys. Lett. A {\bf 242}, 198 (1998).
\bibitem{GKJ16} A. Gily\'{e}n, T. Kiss and I. Jex, Sci. Rep. {\bf6}, 20076 (2016).
\bibitem{TBA+17} J. M. Torres, J. Z. Bern\'{a}d, G. Alber, O. K\'{a}lm\'{a}n and T. Kiss, Phys. Rev. A {\bf95}, 023828 (2017).
\bibitem{MJK+19} M. Malachov, I. Jex, O. K\'alm\'an and T. Kiss, Chaos {\bf 29}, 033107 (2019).
\bibitem{KK18} O. K\'{a}lm\'{a}n and T. Kiss, Phys. Rev. A {\bf97}, 032125 (2018).
\bibitem{I87} I. D. Ivanovic, Phys. Lett. A {\bf123}, 257 (1987).
\bibitem{D88} D. Dieks, Phys. Lett. A {\bf126}, 303 (1988).
\bibitem{P88} A. Peres, Phys. Lett. A {\bf128}, 19 (1988).
\bibitem{HMG+96} B. Huttner, A. Muller, J. D. Gautier, H. Zbinden and N. Gisin, Phys. Rev. A {\bf 54}, 3783-3789 (1996).
\bibitem{DB02} M. Du\u{s}ek and V. Bu\v{z}ek, Phys. Rev. A {\bf 66}, 022112 (2002).
\bibitem{BCS+08} L. Bartu\v{s}kov\'{a}, A. \v{C}ernoch, J. Soubusta and M. Du\v{s}ek, Phys. Rev. A {\bf 77}, 034306 (2008).
\bibitem{SPD+16} M. A. Sol\'{i}s-Prosser, A. Delgado, O. Jimenez, and L. Neves, Phys. Rev. A {\bf 93}, 012337 (2016).
\bibitem{LB99} S. Lloyd and S. L. Braunstein, Phys. Rev. Lett. {\bf 82}, 1784 (1999).
\bibitem{ACM02} G. M. D'Ariano, C. Macchiavello and L. Maccone, Fortschr. Phys. {\bf 48}, 573 (2000).
\bibitem{CN02} I. L. Chuang and M. A. Nielsen, Quantum Computation and Quantum Information, Cambridge University Press (2002).
\bibitem{KB00} P. Kok and S.L. Braunstein, Phys. Rev. A {\bf62}, 064301 (2000).
\bibitem{M06} J. W. Milnor,  {\it Dynamics in One Complex Variable},  Annals of Mathematical Studies (Princeton University Press, 2006).
\bibitem{ML93} J. W. Milnor and T. Lei, Experiment. Math. {\bf 2}, 37 (1993).
\bibitem{XZQ+15} P. Xue, R. Zhang, H. Qin, X. Zhan, Z. H. Bian, J. Li and B. C. Sanders, Phys. Rev. Lett. {\bf114}, 140502 (2015).
\bibitem{BLQ+15} Z. H. Bian, J. Li, H. Qin, X. Zhan, R. Zhang, B. C. Sanders and P. Xue, Phys. Rev. Lett. {\bf 114}, 203602 (2015).
\bibitem{ZZL+16} X. Zhan, X. Zhang, J. Li, Y. S. Zhang, B. C. Sanders and P. Xue, Phys. Rev. Lett. {\bf116}, 090401 (2016).
\bibitem{ZKK+17} X. Zhan, P. Kurzynski, D. Kaszlikowski, K. K. Wang, Z. H. Bian, Y. S. Zhang and P. Xue, Phys. Rev. Lett. {\bf119}, 220403 (2017).
\bibitem{ZCL+17} X. Zhan, E. G. Cavalcanti, J. Li, Z. H. Bian, Y. S. Zhang, H. M. Wiseman and P. Xue, Optica {\bf4}, 966-971 (2017).
\bibitem{WEX+18} K. K. Wang, C. Emary, M. Y. Xu, X. Zhan, Z. H. Bian, L. Xiao and P. Xue, Phys. Rev. A {\bf97}, 020101(R) (2018).
\bibitem{WWZ+18} K. K. Wang, X. P. Wang, X. Zhan, Z. H. Bian, J. Li, B. C. Sanders and P. Xue, Phys. Rev. A {\bf97}, 042112 (2018).
\bibitem{LZB+17} L. Xiao, X. Zhan, Z. H. Bian, K. K.Wang, X. Zhang, X. P. Wang, J. Li, K. Mochizuki, D. Kim, N. Kawakami, W. Yi, H. Obuse, B. C. Sanders and P. Xue, Nat. Phys. {\bf13}, 1117 (2017).
\bibitem{ZXB+17} X. Zhan, L. Xiao, Z. H. Bian, K. K. Wang, X. Z. Qiu, B. C. Sanders, W. Yi and P. Xue, Phys. Rev. Lett. {\bf119}, 130501 (2017).
\bibitem{WXQ+18} X. P. Wang, L. Xiao, X. Qiu, K. K. Wang, W. Yi and P. Xue, Phys. Rev. A {\bf98}, 013835 (2018).
\bibitem{XQW+18} L. Xiao, X. Qiu, K. K. Wang, Z. H. Bian, X. Zhan, H. Obuse, B. C. Sanders, W. Yi and P. Xue, Phys. Rev. A {\bf98}, 063847 (2018).
\bibitem{nc19} K. K. Wang, X. Qiu, L. Xiao, X. Zhan, Z. H. Bian, B. C. Sanders, W. Yi and P. Xue, Nat. Commun. {\bf10}, 2293 (2019).
\bibitem{prl19} K. K. Wang, X. Qiu, L. Xiao, X. Zhan, Z. H. Bian, W. Yi and P. Xue, Phys. Rev. Lett. {\bf122}, 020501 (2019).
\bibitem{LS00} S. Lloyd and J.-J. E. Slotine, Phys. Rev. A {\bf62}, 012307 (2000).
\end{references}
\end{document}